\documentclass[review,number,sort&compress,3p]{elsarticle}

\usepackage{amssymb}
\usepackage{amsmath}
\usepackage{amsbsy}
\usepackage{subfigure}
\usepackage{graphicx}
\usepackage{hyperref}
\usepackage{multirow}

\begin{document}

\begin{frontmatter}

\title{Three topologies of deep neural networks for pulse height extraction}

\author[srg]{Alberto Regad\'io\corref{c1}}
\ead{regadioca@inta.es}

\author[srg]{J. Ignacio G. Tejedor}
\ead{ignacio.garcia@uah.es}

\author[ore]{Luis Esteban}
\ead{luis.esteban@oit.edu}

\author[srg]{Sebasti\'an S\'anchez-Prieto}
\ead{sebastian.sanchez@uah.es}

\address[srg]{Space Research Group, Universidad de Alcal\'a, 28805 Alcal\'a de Henares, Spain}

\address[ore]{EERE Department, Oregon Institute of Technology, Campus Dr SE 97601, Klamath Falls, Oregon State University, USA}

\cortext[c1]{Corresponding Author}

\begin{abstract}
Pulse shaping is a common technique for optimizing signal-to-noise ratio (SNR) in particle detectors. Although analog or digital linear shapers are typically used for this purpose, there are nonlinear approaches, such as neural networks (NN), which have demonstrated their potential to outperform linear ones. Their nonlinear nature makes it possible to optimize the SNR of incoming pulses and extract diverse information, such as particle type and energy, with extremely short shaping time to avoid crowding. This paper shows three different NNs for shaping pulses: (a) convolutional NN (CNN); (b) recurrent NN (RNN); (c) self-attenuating NN. These NNs shape the pulses and return them unfolded avoiding stacking, and even estimate the height of the pulses when there has been saturation in the preamplifier. In this work we show the architectures of the NNs and their results using CR--RC pulses with Brownian and white noise, but they could be extrapolated to any shape and type of noise. The results obtained show that when the noise level is low and the frequency is low, all the topologies presented are a valid solution, but with white noise and high pulse arrival frequency, CNN is a better solution than the others. In the case of Brownian noise, the three topologies presented give similar results.

\end{abstract}


\end{frontmatter}

\section{Introduction}
\label{Introduction}

When particles interact with detectors, pulses of current or charge are generated. These pulses are converted to voltage at the output of a preamplifying stage and shaped afterwards. Finally, the shaped pulses are analyzed to extract diverse information, such as particle type and energy. The ideal shape depends on the pulse coming from the preamplifier, the noise spectrum of the entire system. Thus, specific techniques are used to find out shapes to maximize their Signal-to-Noise Ratio (SNR) and synthesize them, all this without losing sight of the ballistic deficit and pile-up \cite{Knoll2010, Nakhostin2017}.

One of the most common type of noise in spectroscopy systems is white noise, its spectrum is the same in all frequencies. The impact of this type of noise in measurements is inversely proportional to the shaping time as claimed by \cite{Radeka1968, Goulding1972}. Thus, a common practice to mitigate the noise is making the pulse longer with the consequent risk of pile-up. However, the pile-up problem is twofold. On the one hand, the pulse processing complexity must be increased to obtain the height of the incident particles contained in the pulses. On the other hand, the shaping stage can become saturated, thus invalidating the height measurements. When pile-up occurs in detectors, many algorithms detect and discard the piled-up pulses with the consequent loss of information. However, there are algorithms that try to analyze the signals even when they are piled-up (e.g. \cite{Imperiale2001, Kafaee2020}).

Apart from specific algorithms, to address this issue, we can use a special case of pulse shaping called the unfolding technique (or deconvolution) that allows the transformation of the digitized signal into a unitary pulse.

Apart from specific algorithms, we can use a special technique of pulse shaping such as the unfolding (also called deconvolution) that allows the transformation of the digitized signal into a unit impulse $\delta[n]$ in the discrete-time domain \cite{Jordanov2016, Regadio2018} avoiding the pile-up and giving the pulse height of each event with need of additional signal processing.

Despite that unfolded pulses is an optimal way to deal with pile-up, unfolding is not frequently used because its short shaping time increases the effect of the white noise, as explained in the previous paragraph. To avoid this problem, we propose to replace the linear shaper with a nonlinear one, capable of splitting the pulses while keeping the white noise low.

However, the design of a non-linear filter can be tedious. To overcome this problem, we have chosen to implement it by means of artificial Neural Networks (NNs) because they have non-linear response and are automatically configured from input and output patterns. In addition, the use of NNs allow to improve the SNR ratio beyond what a linear system allows because they have a non-linear response.

Machine learning, particularly neural networks, has been widely used in the field of physics to the extent that it is considered an additional tool for its development \cite{Workman2022}. One of the reasons for this is that machine learning has a mathematical formulation intimately related to statistics. This link between machine learning and physics has motivated the creation of a review on Machine Learning for Particle and Nuclear Physics.

One of the most commonly used networks in this field has been the perceptron \cite{Rosenblatt1961}, typically in a multilayered version. There are several reasons for this, one of the most important being that, although there are more novel topologies, multilayer perceptrons have proven to be effective in a wide range of prediction and modeling problems. They have been used for decades and have demonstrated their effectiveness in a variety of applications.

Perceptrons are used, among many other physics-related applications, to estimate mass excess in nuclei, which is crucial for understanding properties of atomic nuclei such as nucleon binding energy, nuclear reaction Q-value, and energy threshold \cite{Ozdougan2022}. Curve fittings modeling cross-sectional neutron-induced reaction cross-sections around specific energy values are also being conducted using perceptrons. This has application in the field of fusion reactor technology, as understanding these cross-sections can provide information about nuclear heating and radiation damage resulting from gas formation in such reactors \cite{Ozdougan2023}. Perceptrons are also used to determine the cross-sectional reactions that are induced by high-energy photons and are one of the important types of reactions in nuclear structure studies. Utilizing a multilayer perceptron enables the estimation of the cross-section and comparison with those obtained in other experiments \cite{Akkoyun2023}. Lastly, they are used to predict beta decay energy \cite{Munoz2023} using a multilayer perceptron with supervised learning.

Another common use of perceptrons in Nuclear and Particle physics is as filters. Later, with the development of the Deep Learning, different NN architectures were used depending on the different types of input data: convolutional neural networks (CNN) to process spatial data such as images and Recurrent Neural Netwoks (RNN), Long-Short Term Memories (LSTM) and Gated Recurrent Unit (GRU) to process temporal data \cite{Goodfellow2016}. All of them are limited by their ability to take into account inputs that are too far away in space and/or time. Attention-based Neural Networks (ANNs) \cite{Bahdanau2014, Olah2016}, especially Transformers \cite{Vaswani2017}, emerged as NNs that could theoretically provide an output taking into account all past inputs and not only those that are close (in space or time), albeit increasing the processing time.

In nuclear spectroscopy, NNs allow the identification of particles analyzing the pulse shape \cite{Jimenez2012, Flores2016, Scrimaglio2004, Szadkowski2017}. On the other hand, pulse shaping with NNs have been carried out using attention networks as Transformers \cite{Wang2022}, Autoencoders \cite{Regadio2019b}, convolutional U-nets \cite{Regadio2021}, studying the influence of the pulse shape and other parameters such as resolution in bits of the preamplifier and sampling rate with the aim of saving computing resources \cite{Wang2023}.

Theoretically, there are a few reasons for using ANNs instead CNNs for pulse shaping. Firstly, as mentioned above ANNs are better at handling long-range dependencies than CNNs and RNNs. This is because they can directly model the relationships between values in a whole sequence, while CNN can only model local dependencies. This locality depends on the number of layers and their size. Secondly, ANNs have been shown to be more effective at capturing global features than CNN and RNNs because they are able to attend to all parts of the input simultaneously. Thirdly, ANNs are more efficient in terms of computation because they can process the input in parallel, while CNNs and RNNs need to process the input multiple times \cite{Vaswani2017, Geron2022}.

In this paper, we check these assertions comparing three topologies of NNs: (a) CNNs; (b) recurrent NNs; (c) ANNs. All of them perform unfolding, removing the pile-up issue without losing sight of the SNR. In addition, all the presented NNs, apart from unfolding while increasing the SNR, learn how to approximate the height of the pulse chopped when the preamplifier is saturated. Additionally, a new loss function that delays the unfolded pulse removes the problem of ballistic deficit defined as the loss in pulse height that occurs at the output of a shaping network when the input pulse has a rise time greater than zero due to the charge collection time \cite{Loo1988, Nakhostin2017}. The management of all these issues, mixed with the input noise make the training process more laborious.

Comparisons among topologies will be made using the cost function and how they behave for different noise amplitudes and pulse arrival frequency. Note that the number of parameters of the input pulses, the number of topologies used and their corresponding parameters mean that we have to select a reduced number of them to make the comparisons.

Finally, note that in this article we present a way for synthesizing digital pulse shapes, regardless of the detector type or even the application. Therefore, it is not our intention to discuss the application of these shapes, nor their performance in spectroscopy systems or any other systems that may use them.

The rest of the paper is structured as follows: Section \ref{Test bench} describes the input pulses for the three NN topologies. Section \ref{Loss functions} explains the two loss functions used in this paper to train the NNs. Section \ref{Architectures} explains the principles, architectures, implementation and the results of each NN whereas the comparison among them is shown in Section \ref{Comparison and Discussion}. Finally, Section \ref{Conclusions} summarizes the conclusions of this work.

%%%%%%%%%%%%%%%%%%%%%%%%%%%%%%%%%%%%%%%%%%%%%%%%%%%%%%%%%%%%%%%%%%%%%%%%%%%%%%%

\section{Test bench}
\label{Test bench}

The tested NNs were trained with a sequence of samples with pulses  that mimic the generated in particle detectors. These pulses are generated at random time intervals and that can be piled-up. This sequence of samples was mixed with white and brownian noise that are two of the most common noise types in particle detectors \cite{Goulding1972, Nicholson1974, Knoll2010, Nakhostin2017, Workman2022}. Once trained, it was tested with sequences of samples different from the training dataset, as the basic neural network testing procedure dictates \cite{Geron2022}.

\subsection{Input pulse shape}
\label{Input pulse shape}

\begin{figure}[!ht]
	\centering
	\includegraphics[scale=0.7]{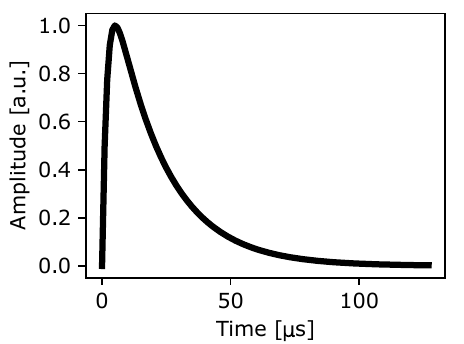}
	\caption{Example of CR--RC pulse.}
	\label{fig:CRRCpulse}
\end{figure}

The proposed NNs can be trained with any input pulse shape. However, throughout the entire article, we will use a CR--RC pulse. The reason is because this shape is typically the output of preamplifiers in particle detectors \cite{Nicholson1974, Knoll2010}. In general, the equation to obtain CR--RC pulses is \cite{Nakhostin2017}
\begin{equation}\label{EqCRRC}
	f(t) = \frac{{\tau}_1}{{\tau}_1 - {\tau}_2} \left( e^{\frac{-t}{{\tau}_1}} - e^{\frac{-t}{{\tau}_2}} \right)
\end{equation}
For this work and for the sake of simplicity, we set $\tau1=1$ $\mu$s, $\tau2=0.1$ $\mu$s and the signal was sampled at $T_s=50$ ns to give a CR--RC pulse as the one depicted in Figure \ref{fig:CRRCpulse}. 

\subsection{Pulse generation}
\label{Pulse generation}

\begin{figure}[!ht]
	\centering
	\includegraphics[scale=0.7]{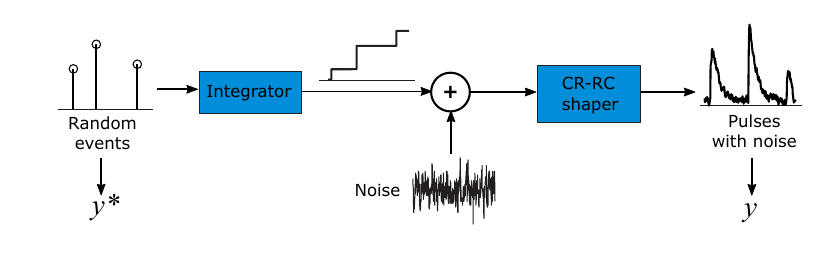}
	\caption{Generation of the input pulses and configuration of the simulation according to the method presented in \cite{Regadio2021}).}
	\label{fig:SimBank}
\end{figure}

Once defined the pulse shape, we used the procedure followed in \cite{Regadio2021} to generate the pulses and the samples that contain them. That is, we used as source step (Heaviside) pulses whose pulse height is a random uniform distribution between 0 and 1 to mimic the pulses generated at the particle detector when the charge is collected. A complete configuration of the simulation is shown in Figure \ref{fig:SimBank}.

To efficiently train and test the NNs, pulses were generated at random time intervals and some of them are piled-up. This can give rise to overlaps that saturate the signal when it is greater than 1. However, even if the pulse is chopped, the network must estimate the height of the pulse, as mentioned in Section \ref{Introduction}. These pulses were mixed with brownian and white noise created with a random procedure. Afterwards, the noisy pulses were shaped to CR--RC ones introduced in Section \ref{Input pulse shape}. As specified in Section \ref{Input pulse shape}, the sampling frequency will be fixed and equal to 20 MHz ($T_s$ = 50 ns).

We set the sample length equal to 1024 in order to be able to observe clearly the noise effects and the pile-up without overloading the training time of the neural network, especially the RNN.

\begin{figure}[!ht]
	\centering
	\includegraphics[scale=0.7]{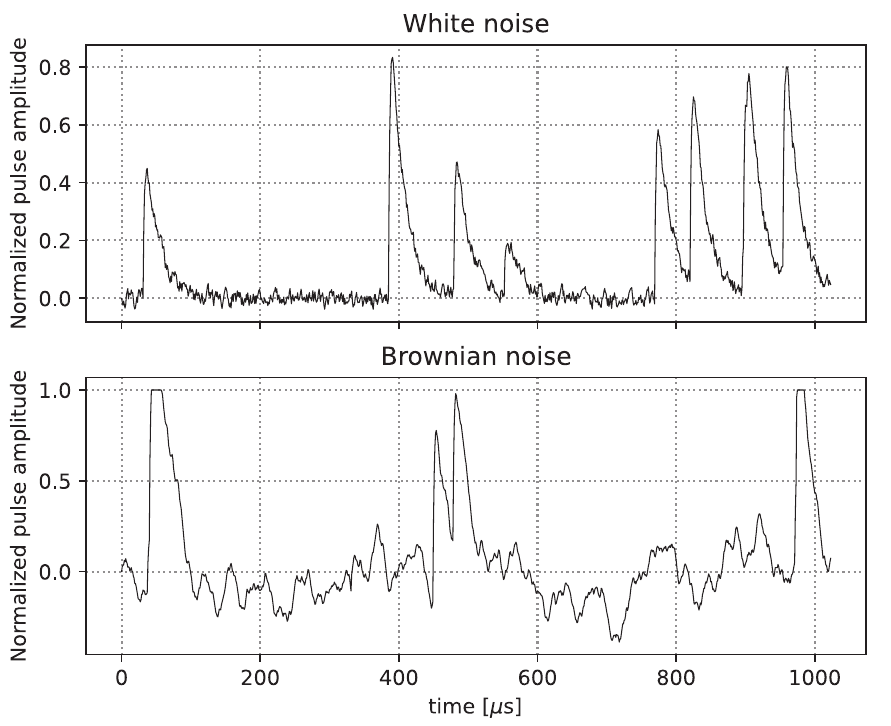}
	\caption{Example of CR--RC pulses mixed with white noise (top panel) and brownian (bottom panel).}
	\label{fig:ExampleOfPulses}
\end{figure}

The tested pulse mean count rate varies from $8 \cdot {10}^{3}$ to ${1.6 \cdot 10}^{2}$ s${}^{-1}$. A sample of train of pulses with white noise amplitude equal to 0.1 and pulse mean count rate equal to $1.6 \cdot {10}^{2}$ s${}^{-1}$ is shown in the top panel of Figure \ref{fig:ExampleOfPulses}. Another sample of train of pulses with brownian noise amplitude equal to 0.1 and pulse mean count rate equal to $1.6 \cdot {10}^{2}$ s${}^{-1}$ is shown in the top panel of Figure \ref{fig:ExampleOfPulses}.

During the training process, these noisy CR--RC pulses were used as input $\mathbf{x}$ of the proposed NNs whereas unfolded pulses without noise were used at the output $\mathbf{y^{\ast}}$. However, before training the network, is mandatory to set its parameters.

The procedure described in this Section was entirely programmed in Python using the Numpy library \cite{Harris2020}.

%%%%%%%%%%%%%%%%%%%%%%%%%%%%%%%%%%%%%%%%%%%%%%%%%%%%%%%%%%%%%%%%%%%%%%%%%%%%%%%%%%%%

\section{Loss functions}
\label{Loss functions}

The NNs were trained in a supervised way to convert the output $\mathbf{y}$ into a desired output $\mathbf{y^{\ast}}$. As explained in the previous Section, $\mathbf{y^{\ast}}$ are the unfolded pulses, that is Dirac delta pulses whose amplitude is proportional to the height of the incoming pulse as explained in Section \ref{Introduction}. With the purpose of training our NNs, we used two loss functions $J$, the last one specifically created for pulse height analysis.

\subsection{Mean squared error (MSE)}
\label{Mean squared error}

The Mean Squared Error (MSE) is widely used for training NNs. It is defined as 

\begin{equation}\label{EqMSELoss}
	J = \text{mean} \left( {\left( \mathbf{y}^{\ast}[n] - \mathbf{y}[n] \right)}^2 \right)%{(\mathbf{y^{\ast}}-\mathbf{y})}^2
\end{equation}

We have used the implementation of this equation provided by Tensorflow \cite{TENSORFLOW2015}. 

\subsection{Multiple mean squared error (MMSE)}
\label{Multiple mean squared error (MMSE)}

The MSE function is widely used. However, in nuclear spectroscopy, when extracting the height of incoming pulses it generally does not matter if these heights arrive with a reasonable delay when the calculated heights are more accurate. This is particularly true when we can have ballistic deficit as discussed in Section \ref{Introduction}.

For this purpose we have created this loss function. With it, the neural network automatically chooses what delay of unfolded pulses yields more accurate heights. The output is therefore transformed from $\delta[n]$ to $\delta[n-k]$. The new function is:

\begin{equation}\label{EqMultipleMSELoss}
	J =  \min_{k} \left( \text{mean} \left( {\left( \mathbf{y}^{\ast}[n-k] - \mathbf{y}[n] \right)}^2 \right) \right)%{(\mathbf{y^{\ast}}-\mathbf{y})}^2
\end{equation}
where $k \in [0, K]$. For simplicity, we set $K=5$ along the entire article. We implemented this function in Tensorflow \cite{TENSORFLOW2015} deriving the \textit{tf.keras.losses} class. 

%%%%%%%%%%%%%%%%%%%%%%%%%%%%%%%%%%%%%%%%%%%%%%%%%%%%%%%%%%%%%%%%%%%%%%%%%%%%%%%%%%%%

\section{Architectures}
\label{Architectures}

Once established the dataset and the loss functions, we move on to the most complex part, which is to choose the NNs that will act as a filter for each topology\footnote{Data and code is available on demand.}.

When dealing with a problem involving NNs and signal processing, two options are typically chosen: (a) generate a spectrogram of the pulse and process it as if the signal were a two-dimensional image; (b) process the signal as a temporal series (one dimension). In this article we are going to focus in the latter approach to avoid transformations that, in principle, should be carried out automatically by NNs.

The entire implementation of the proposed NNs was programmed in Python using Tensorflow \cite{TENSORFLOW2015} and Keras \cite{KERAS2015} packages.

\subsection{Convolutional NN}
\label{Convolutional NN}

Convolutional neural networks (CNNs) are a specific topology of NNs well suited for processing spatial data with structure in one, two or more dimensions, such as images. They are also used for other types of data represented as sequences or time series. In particular, CNNs are commonly used for tasks such as image and audio recognition, denoising, and compression. Overall, CNNs have proven to be a powerful and versatile tool for processing data with a grid-like structure, making them an important tool in the field of signal processing.

CNNs consist of several layers, including convolutional layers, pooling layers, and fully connected layers, and even recurrent layers as LSTM or GRU layers when processing temporal data \cite{Goodfellow2016, Geron2022}. Each layer is designed to extract and process different features of the input data, allowing the CNN to learn increasingly complex representations of the data.

\begin{figure}[!ht]
	\centering
	\includegraphics[scale=0.8]{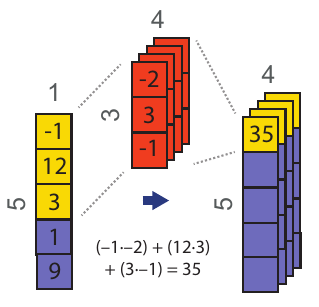}
	\caption{Example of 1D convolution layer of size 3.}
	\label{fig:ConvoExample}
\end{figure}

\begin{figure}[!ht]
	\centering
	\includegraphics[scale=0.8]{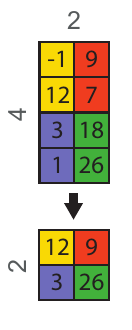}
	\caption{Example of max pooling layer of size 2.}
	\label{fig:MaxPoolExample}
\end{figure}

Convolutional layers are the core of CNNs and work by applying a set of filters to the input data. These filters can be thought of as small windows that slide over the input data, computing a dot product between the filter and the input at each position (i.e., a convolution) \ref{fig:ConvoExample}. The result of this operation is a feature map that highlights the presence of certain patterns in the input data. Pooling layers are often used after convolutional layers to reduce the dimensionality of feature maps, making the CNN more efficient and robust to small variations in the input data \cite[p. 330]{Goodfellow2016}. In this section, we used max pooling. It works by partitioning the input into a set of non-overlapping rectangles and, for each such sub-region, outputs the maximum value \ref{fig:MaxPoolExample}. This helps in retaining the most important features while reducing the computational load and controlling overfitting.

\begin{figure}[!ht]
	\centering
	\includegraphics[scale=0.6]{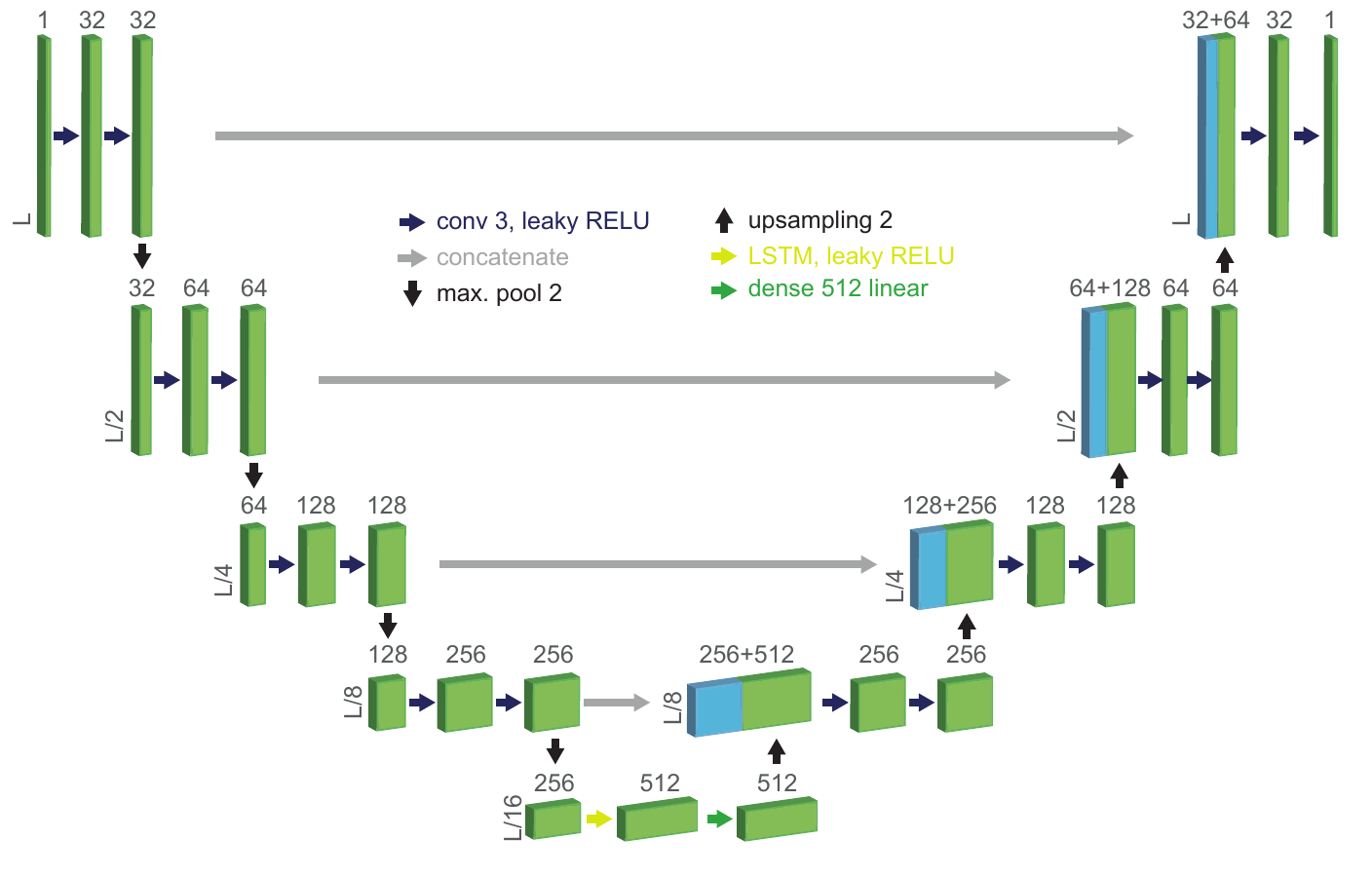}
	\caption{U-net architecture presented in \cite{Regadio2021}.}
	\label{fig:ArchitectureUnet}
\end{figure}

U-Net is an specific type of CNN applied to a wide range of image processing applications. U-Net gets its name from its shape, which resembles a ``U'' (Figure \ref{fig:ArchitectureUnet}). The architecture of U-Nets consist of two main parts: the contracting path and the expansive path. The contracting path is similar to a traditional CNN, consisting of several convolutional and pooling layers that progressively down-sample the input data. The purpose of this path is to extract increasingly abstract features from the input data.

The expansive path, on the other hand, consists of several up-sampling and convolutional layers that progressively up-sample the feature maps from the contracting path. The purpose of this path is to reconstruct the output image or segmentation mask by gradually adding more and more detail back into the output. The final layer of the network is usually a softmax layer that outputs a probability map indicating the likelihood of each pixel belonging to a certain class but, in \cite{Regadio2021}, the output is the unfolded signal.

In Section \ref{Comparison and Discussion}, the results obtained in \cite{Regadio2021} with the same U-net topology of depicted in Figure \ref{fig:ArchitectureUnet} are going to be compared with the topologies exposed in this article, with the novelty that we will also use the cost function of Eq. (\ref{EqMultipleMSELoss}) apart from MSE (\ref{EqMultipleMSELoss}). For examples of processing and additional information about this NN, see the article reference.

\subsection{Recurrent NN}
\label{Recurrent NN}

Recurrent neural networks (RNNs) are a type of neural network that is well-suited for processing temporal data. Unlike feedforward neural networks such as \cite{Rosenblatt1961}, which process input data in a fixed order, RNNs have loops that allow information to be passed from one time step to the next. This allows RNNs to maintain a kind of ``memory'' of previous inputs \cite{Goodfellow2016, Geron2022}, making them particularly useful for processing time-series.

In RNNs, each time step has its own set of weights and biases, which are shared across all time steps. The input to the network at each time step is combined with the output from the previous time step to produce the current output. This allows the network to capture dependencies between input data at different time steps. However, traditional RNNs can suffer from the problem of vanishing gradients, where the gradients used for training the network become very small and the network stops learning long-term dependencies. Long Short-Term Memory (LSTM) and Gated Recurrent Units (GRU) were developed as solutions to this issue.

LSTM networks can learn long-term dependencies by selectively ``forgetting'' or ``remembering'' information from previous time steps. LSTMs have a memory cell that can store information for a long period of time, as well as several ``gates'' that control the flow of information into and out of the cell. These gates allow LSTMs to selectively forget or remember information from previous time steps, enabling them to capture long-term dependencies in the input data.

The Gated Recurrent Unit (GRU) networks can learn long-term dependencies by selectively updating their internal state based on the current input and the previous state. GRUs also have ``gates'' that control the flow of information, but they have a simpler architecture than LSTMs, with fewer parameters to learn. It works by using the update gate $z_t$ to determine how much of the previous memory content $h_{t-1}$ should be retained and how much of the new memory content $\tilde{h}_t$ should be integrated. The reset gate $r_t$ helps the model decide how much of the previous memory content should be ignored when computing the new memory content. The new memory content $\tilde{h}_t$ is a combination of the reset gate's influence on the previous memory and the input at the current time step. Finally, the final memory content $h_t$ is a combination of the previous memory content scaled by $1 - z_t$ and the new memory content scaled by $z_t$. This mechanism allows the GRU to capture long-term dependencies and address the vanishing gradient problem often encountered in traditional RNNs. GRUs have been shown to be effective in a wide range of tasks, including signal processing. Its network specifications are extensively detailed in the literature \cite[p. 413]{Goodfellow2016}, \cite{Geron2022}. However, we list below the formulae of its implementation in Keras for this project. The equation of the update gate $z_t$ is the following:
\begin{equation}
z_t = \sigma(W_z \cdot [h_{t-1}, x_t] + b_z)
\end{equation}

The reset gate is controlled according to this equation:
\begin{equation}
r_t = \sigma(W_r \cdot [h_{t-1}, x_t] + b_r)
\end{equation}

New memory content, that refers to the candidate activation that could be added to the memory, is modelled with this formula:
\begin{equation}
\tilde{h}_t = \text{tanh}(W_h \cdot [r_t \cdot h_{t-1}, x_t] + b_h)
\end{equation}

Finally, the updated memory content is controlled with:
\begin{equation}
h_t = (1 - z_t) \cdot h_{t-1} + z_t \cdot \tilde{h}_t
\end{equation}

In all of these equations, $x_t$ is the input at time step $t$, $h_t$ is the output at time step $t$, $W_z$, $W_r$, and $W_h$ are weight matrices, $b_z$, $b_r$, and $b_h$ are bias vectors, and $\sigma$ is the sigmoid function.

\begin{figure}[!ht]
	\centering
	\includegraphics[scale=0.8]{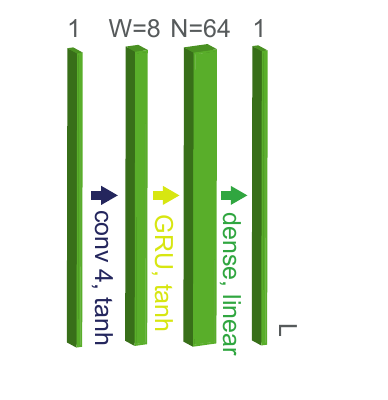}
	\caption{Proposed recurrent architecture.}
	\label{fig:ArchitectureRNN}
\end{figure}

In this article we have opted for the latter because previous tests have shown a better result with GRUs. Concretely, in this work, we propose the topology depicted in Figure \ref{fig:ArchitectureRNN}. It process the dataset exposed in Section \ref{Test bench} in the same one that the presented in Section \ref{Convolutional NN}.

\begin{figure}[!ht]
	\centering
	\includegraphics[scale=0.5]{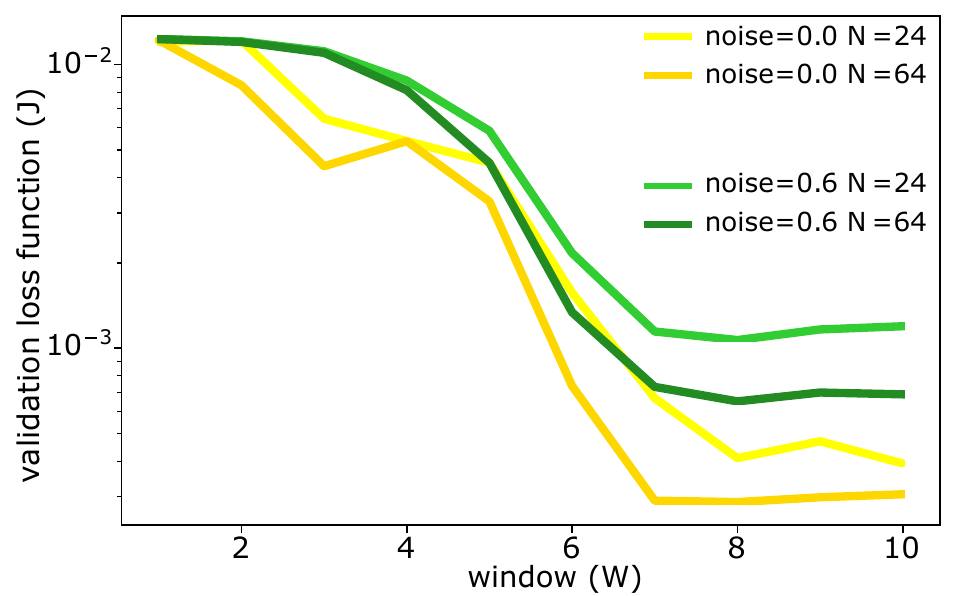}
	\caption{Loss function $J$ for different values of $W$, $N$ and white noise levels.}
	\label{fig:calibration}
\end{figure}

In pursuance of obtaining good results and avoiding overfitting, this RNN was trained with different values of $W$ and $N$ (see Figure \ref{fig:ArchitectureRNN}). Using the loss function $J$ of Eq. \ref{EqMSELoss} and different white noise levels the results are shown in Figure \ref{fig:calibration}. White noise has been chosen because is the most common in particle detectors as stated in Section \ref{Introduction}.

We observe that regardless of the number of neurons used and the noise level, the loss function falls to minimum values that are reached and maintained when $W \ge 8$ and $N=64$.

Furthermore, we observe that, obviously, the higher the value of $N$, the lower the cost function. For all these reasons and with the aim of using the minimum number of neurons to avoid overfitting and not using brute force to see the potential of this architecture, we chose $W=8$ and $N=64$ that yield an NN with 9569 parameters. This number of parameters implies that it can also be inserted in an embedded computer with enough memory and carry out the processing on-line.

After setting the dataset (Section \ref{Test bench}), the loss function $J$ (Section \ref{Loss functions}), and the NN parameters, the next step is training it. A summary of the training results is shown in Table \ref{tab:Comparison}.

\begin{figure}[!ht]
	\centering
	\includegraphics[scale=0.8]{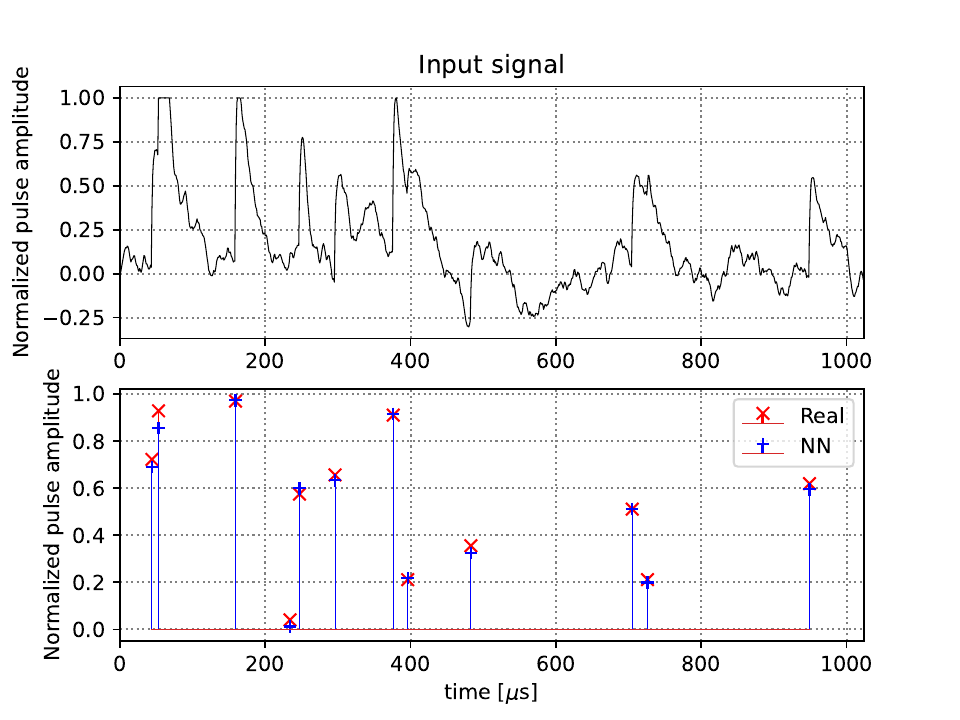}
	\caption{Example of pulse height extraction of CR--RC pulses mixed with brownian noise.}
	\label{fig:ExampleGRU1}
\end{figure}

In order to check the performance of this RNN, a set of preliminary simulations in time-domain were carried out using different noise amplitudes, noise types and pulse rates. An example of such results is shown in Figure \ref{fig:ExampleGRU1} where the brownian noise amplitude is equal to 0.1 and pulse mean count rate equal to $1.6 \cdot {10}^{-2}$ s${}^{-1}$. Note that the pulse height is detected even when the signals are piled-up. It even approximates the height of the pulse when, due to pile-up, the input signal is saturated. The threshold of the pulses was set to 0.003.

Apart from this example, detailed information about this RNN will be exposed in Section \ref{Comparison and Discussion}.

\subsection{Sequence-to-sequence self-attention NN}
\label{Seq2seq self-attention NN}

Sequence-to-sequence (seq2seq) models with attention are a variant of seq2seq models \cite{Sutskever2014, Cho2014} (also known as encoder-decoder) that use only recurrence and use an attention mechanism (\cite{Bahdanau2016}, for this section). 

The attention mechanism is a concept enables models to selectively focus on specific parts of input data, assigning different importance weights to these parts. It allows the model to emphasize relevant information while filtering out noise or irrelevant details. This mechanism is used in improving the performance of seq2seq models in tasks such as machine translation or text generation. They have been traditionally used to enhance the translation or text generation process (see \cite{Tiwari2020} among others).

The attention mechanism is typically implemented in deep learning models through a series of mathematical operations. One common approach is the use of dot product attention, where the input is transformed into query $Q$, key $K$, and value $V$ vectors. These vectors are then used to compute attention scores, which are subsequently converted into attention weights using a softmax function. Finally, the weighted sum of the values is computed to obtain the final output. This is summed up in the following equation \cite{Vaswani2017}:

\begin{equation}
\text{Attention}(Q,K,V) = \text{softmax}\left(\frac{QK^T}{\sqrt{d_k}}V\right)
\end{equation}
where $d_k$ denotes the dimension of the key matrix and ``softmax'' is a mathematical function that takes as input a vector of real numbers and normalizes it into a probability distribution, where the sum of the probabilities is equal to 1.

Like other NNs, attention layers can contain more than one layer (multi-attention layer). Besides, it is possible to use techniques such as residualization and layered normalization \cite{Geron2022} to improve its efficiency.

In general, seq2seq models with attention are effective in learning long-range dependencies in sequential data and, therefore, can be useful for processing signals in one dimension with complex structures that is the objective of this paper.

Attention and self-attention are two related but different techniques commonly used in NNs. Attention refers to the ability of a model to focus on specific parts of the input at each processing step based on the output already determined by the network, while self-attention is a particular form of attention where the relevance between different parts of the input is calculated. To determine when to use an attention layer versus a self-attention layer, it is necessary to consider the specific task. In this paper we chose self-attention instead attention and transformers because we have observed that the output $y[n]$ sometimes becomes unstable with very small variations of past outputs $y[n-1], y[n-2], \ldots$. However, this does not mean that attention is not valid for shaping other pulse shaping tasks as has already been shown in \cite{Wang2022}.

\begin{figure}[!ht]
	\centering
	\includegraphics[scale=0.8]{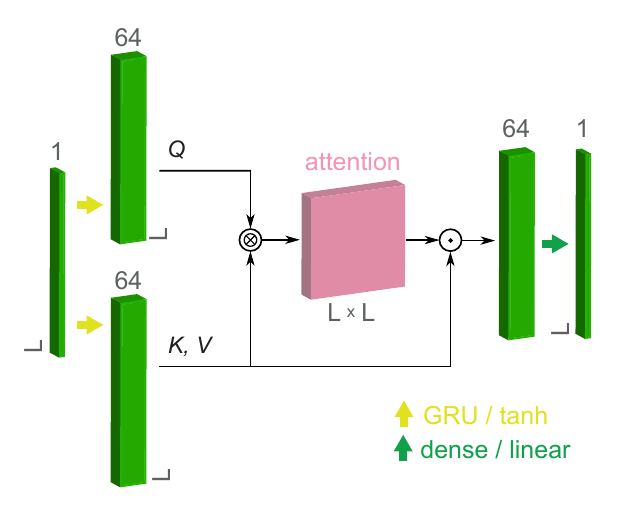}
	\caption{Scheme of the attention mechanism applied to pulse detection. The size of the window is $L$. Blue boxes contain trainable parameters. $\otimes$ denotes tensor product and $\cdot$ stands for dot product.}
	\label{fig:ArchitectureSelfAtt}
\end{figure}

With seq2seq with windowed self-attention, we make sure that the input signal is not forgotten when the time goes by, as it occurs with a bare RNN, as explained in Section \ref{Introduction}. A complete scheme of the proposed self-Attentional Neural Network (ANN) is depicted in Figure \ref{fig:ArchitectureSelfAtt}. The input is processed with two GRU layers to generate the $Q$, $K$ and $V$ vectors. In compliance with to \cite{Vaswani2017}, where they go deeper into their definition, the key and query are used to attend to the input via outer product $\otimes$ whereas the value is used to compute a value whose internal product with trainable parameters gives the output. For the ANN proposed, $K$, and $V$ are the same due to a trade-off of efficiency and performance.

Once $Q \otimes K$ is calculated to yield a matrix whose entries to calculate how much $Q$ and $K$ match, this matrix is multiplied by $V$. This value will be passed through another dense layer (i.e. calculate the inner product between two vectors) to give a value that is the desired value at time $t$. 

\begin{figure}[!ht]
	\centering
	\includegraphics[scale=0.8]{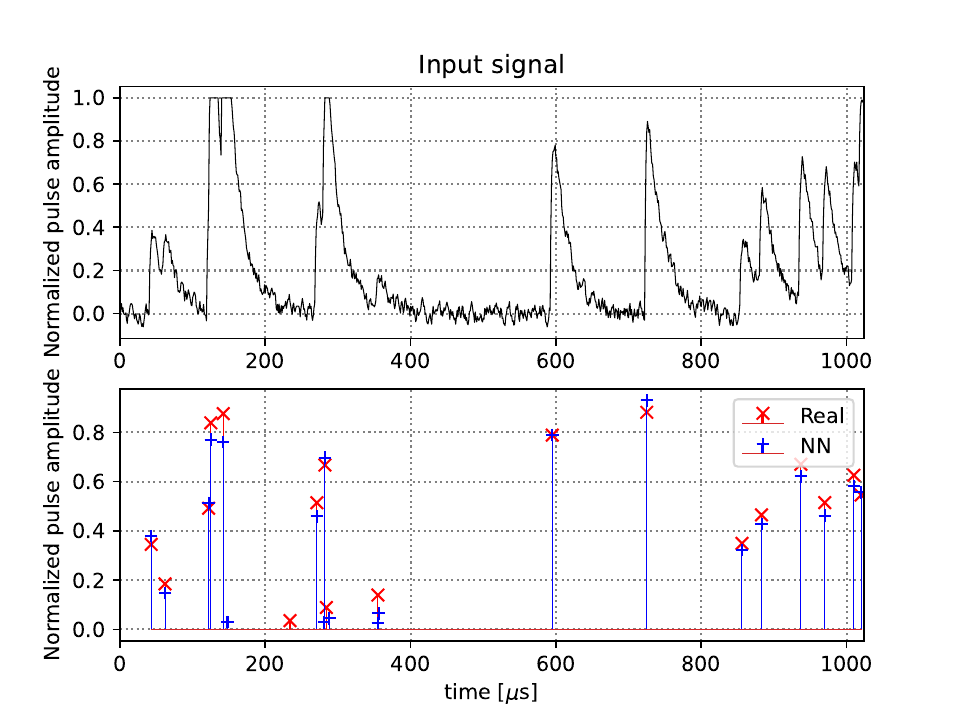}
	\caption{Example of pulse height extraction of CR--RC pulses mixed with white noise.}
	\label{fig:ExampleSelfAtt0}
\end{figure}

In the same way that in Section \ref{Recurrent NN}, to check the performance of this ANN, a set of preliminary simulations in time-domain were carried out using different noise amplitudes, noise types and pulse rates. An example of such results is shown in Figure \ref{fig:ExampleSelfAtt0} where we see an example of this NN filtering white noise amplitude is equal to 0.1 and pulse mean count rate equal to $1.6 \cdot {10}^{-2}$ s${}^{-1}$. Note that, in the same way that the CNN and the RNN the pulse height is detected even when the signals are piled-up. It even approximates the height of the pulse when, due to pile-up, the input signal is saturated. The threshold of the pulses was set to 0.003.

Apart from this example, detailed information about this NN in comparison with the other ones will be exposed in Section \ref{Comparison and Discussion}.

%%%%%%%%%%%%%%%%%%%%%%%%%%%%%%%%%%%%%%%%%%%%%%%%%%%%%%%%%%%%%%%%%%%%%%%%%%%%%%%

\section{Comparison and Discussion}
\label{Comparison and Discussion}

Each NN type presented in previous Section have been trained with 1800 training pulses and 800 validation pulses. The loss functions presented in Section \ref{Loss functions} and the Adam optimizer at a learning rate equal to 0.0005. Every training converge. The results of the training are summarized in Table \ref{tab:Comparison}.

\begin{table}[htb]
	\centering
		\begin{tabular}{lcrcccc}
			\hline
			&&&&&\multicolumn{2}{c}{Loss ($\times 10^{-4}$)} \\
			Type & L & Parameters & Epochs & Training time & MSE & MMSE \\
			\hline
			Convolutional          & 1024 & 3274305 & 20 & 78 s/epoch & 2.89 & 0.30 \\
			Recurrent              & 1024 & 9569    & 25 & 32 s/epoch & 4.37 & 0.48 \\
			Seq2seq auto-attention & 1024 & 6981    & 25 & 98 s/epoch & 0.81 & 0.80 \\
			\hline
		\end{tabular}
	\caption{Comparison between number of parameters (neurons), learning parameters and achieved loss function ($J$) of the three topologies used.}
	\label{tab:Comparison}
\end{table}

As the sequences have been generated by simulation, we can obtain as many as we want to improve the training of the ANN at the cost of increasing the training time. However, an additional application of this filter may be to regenerate the shaper when the features of the detector have changed as a consequence of radiation, as for example in \cite{Lanchares2013} as it can be the case with silicon detectors installed on payloads. Therefore for these experiments a tradeoff between training time and performance was chosen.

We can observe that, despite having far fewer parameters, when we use the MMSE loss function (Section \ref{Multiple mean squared error (MMSE)}), the RNN is similar in performance to the CNN with far fewer parameters. Furthermore, with this loss function, the RNN outperforms the ANN.

The ANN, despite the advantages explained in Section \ref{Seq2seq self-attention NN}, has the disadvantage that the attention mechanism adds additional complexity to the NN. Moreover, in this type of applications, attention has a limited advantage because for pulse processing in principle it is only necessary to keep attention on the last inputs taken; this is something that RNNs carry out perfectly. The RNN is also the fastest to be trained as it has the lowest complexity.

Apart from this previous discussion, to verify in a more quantitative way how the network behaves against noise, we have compared, by means of two-dimensional histograms, the difference between the heights of the real pulses and those calculated by the NNs. But firstly, we will evaluate how two classical filterings behave when we introduce frequency and noise levels. In all cases the number of samples is 1000 and networks were trained with the MMSE loss function (Section \ref{Multiple mean squared error (MMSE)}) because it gives better results than MSE according to Table \ref{tab:Comparison}.

\begin{figure}[!ht]
	\centering
	\includegraphics[scale=0.65]{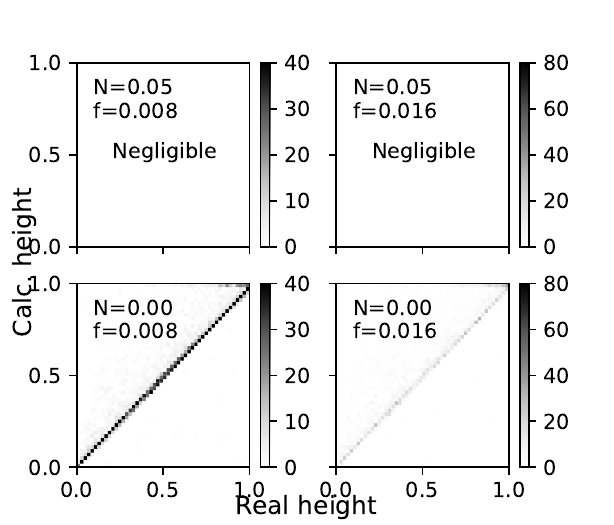}
	\caption{Relationship between the real pulse height and the calculated using the FIR filter $h(z)=\frac{1}{5}(1+z^{-1}+z^{-2}+z^{-3}+z^{-4})$ with different levels of white noise $N$ and pulse frequency $f$.}
	\label{fig:Noise_filt_hist}
\end{figure}

First let's use a low-pass FIR filter of order 5 whose response function in z-domain is $h(z)=\frac{1}{5}(1+z^{-1}+z^{-2}+z^{-3}+z^{-4})$. These type of digital filters are commonly used for nuclear spectroscopy and pulse height analysis \cite{Knoll2010, Nakhostin2017}. We see that for white noise levels, it barely captures the pulse correctly (Figure \ref{fig:Noise_filt_hist}). Note also that, as discussed in Section \ref{Introduction}, there is saturation in the detectors despite the absence of noise. This contributes to the fact that even in the complete absence of noise, it is impossible for this filter to find the height. This same experiment gives similar results with brownian noise.

\begin{figure}[!ht]
	\centering
	\includegraphics[scale=0.65]{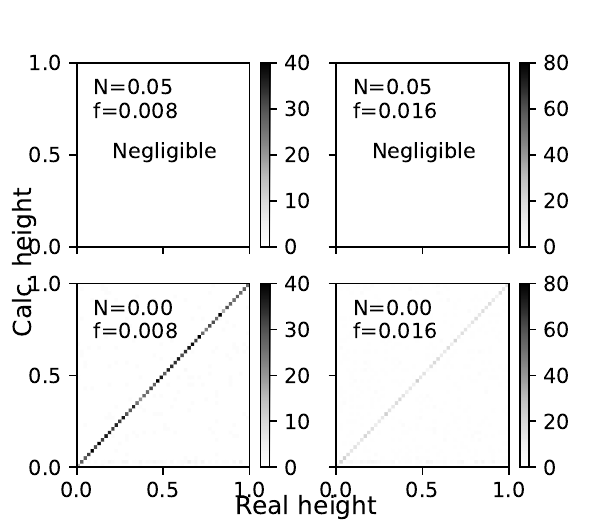}
	\caption{Relationship between the real pulse height and the calculated using unfolding with different levels of white noise $N$ and pulse frequency $f$.}
	\label{fig:Noise_unfo_hist}
\end{figure}

Something similar happens with the second type of filter: an unfolder \cite{Jordanov2016}, as shown in \cite{Regadio2018}. We see that in the absence of noise, the unfolder works perfectly (Figure \ref{fig:Noise_unfo_hist}, in fact much better than the proposed NNs as will be seen later. However, when adding white its performance drops dramatically as also shown in this same figure (in fact, the detected pulses are negligible). As the frequency of the pulses increases, saturation occurs in the detector and it does not detect it either as can be seen in the bottom right panel of the figure \ref{fig:Noise_unfo_hist}. This unfolder yields similar results with brownian noise.

Once we have reference values, we have to do the same test with the three topologies of NN presented.

\begin{figure}[!ht]
	\centering
	\includegraphics[scale=0.65]{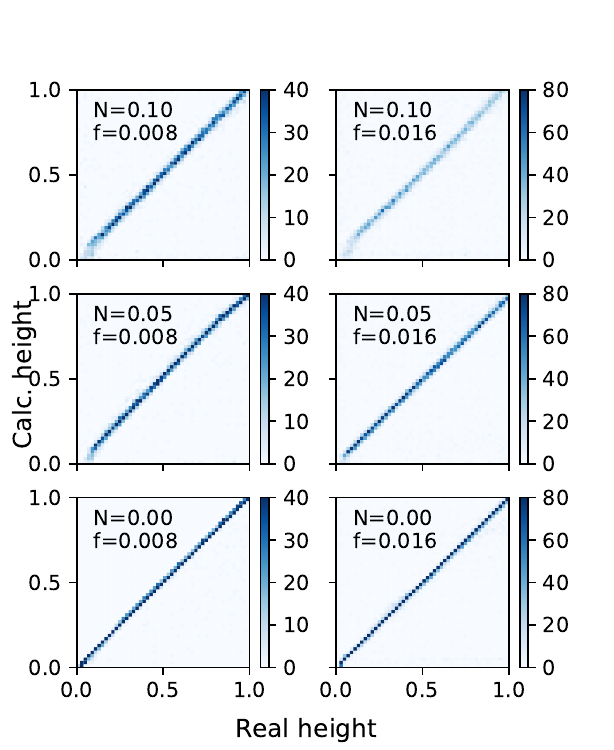}
	\caption{Relationship between the real pulse height and the calculated using the U-net for different levels of white noise $N$ and pulse frequency $f$.}
	\label{fig:Convo_0}
\end{figure}

\begin{figure}[!ht]
	\centering
	\includegraphics[scale=0.65]{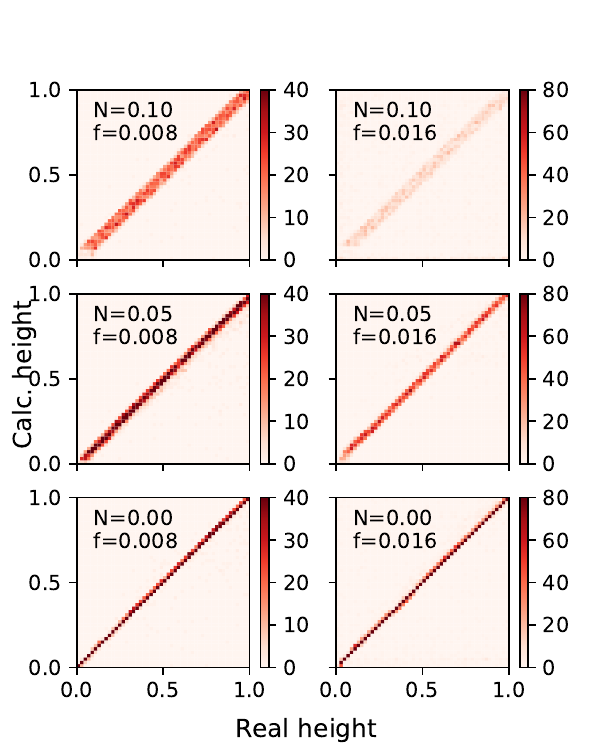}
	\caption{Relationship between the real pulse height and the calculated using the U-net for different levels of brownian noise $N$ and pulse frequency $f$.}
	\label{fig:Convo_1}
\end{figure}

With respect to the CNN, Figure \ref{fig:Convo_0} shows that the correlation between real pulse heights and those calculated by the CNN in presence of white noise are very high for low noise levels and low frequency and that only when the noise and frequency are high (upper panel of Figure \ref{fig:Convo_0}), the CNN starts to fail. In Figure \ref{fig:Convo_1} we can see that brownian noise has more effect than white noise, since when it is high, the correlation line becomes fuzzy. In addition, when the pulse arrival frequency is high, the pulses start to get lost (top right panel of Figure Figure \ref{fig:Convo_1}). To visualize the magnitude of noise levels and pulse frequency, the bottom right panel of figure \ref{fig:Convo_0} corresponds to the noise and frequency of the top panel of Figure \ref{fig:ExampleSelfAtt0} while the top left panel of figure \ref{fig:Convo_1} corresponds the top panel to Figure \ref{fig:ExampleGRU1}.

%In figure \ref{fig:Convo_0} we can see that when the noise has an amplitude of 0.1, the CNN calculates the heights of the pulses with hardly any error. The fact that the pulse arrival frequency and therefore the stacking interfere more in the pulse measurement than the noise itself is also observed. This is also the case since the stacked pulses saturate the detector and the network also has to estimate the height of these pulses.

%In figure \ref{fig:Convo_-1} we can see that brownian noise has more effect than white noise. However, as in the case of white noise, the stacking of pulses still has more effect than the noise itself. It can also be seen that when the stacking is low, there are pulses whose pulse height is estimated to be very low. It is also observed that as the pulse ratio and brownian noise increase, the difficulty of calculating the height of these pulses increases. This is so even if the network parameters $W$ and $N$ are increased.

\begin{figure}[!ht]
	\centering
	\includegraphics[scale=0.65]{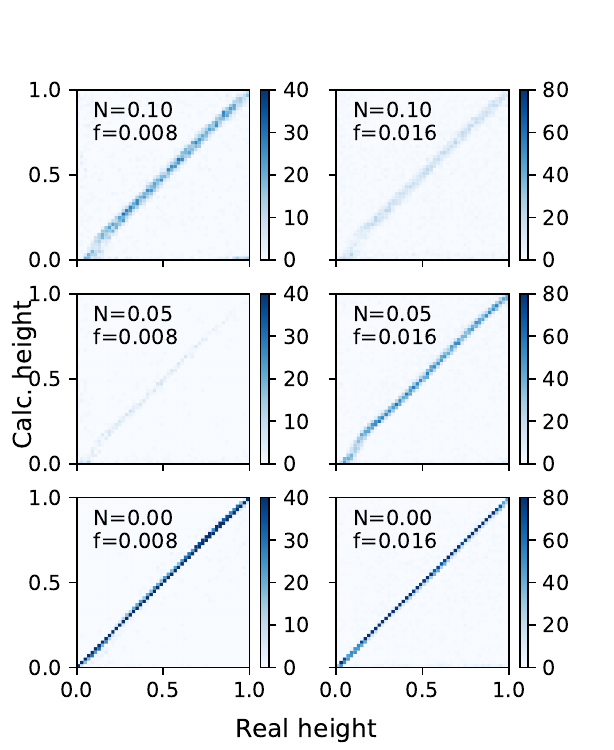}
	\caption{Relationship between the real pulse height and the calculated using the proposed RNN for different levels of white noise $N$ and pulse frequency $f$.}
	\label{fig:Gru_0}
\end{figure}

\begin{figure}[!ht]
	\centering
	\includegraphics[scale=0.65]{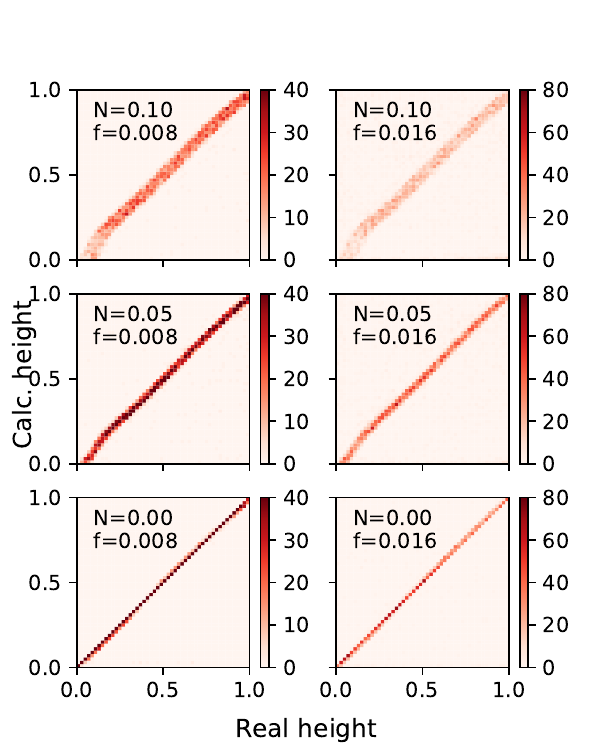}
	\caption{Relationship between the real pulse height and the calculated using the proposed RNN for different levels of brownian noise $N$ and pulse frequency $f$.}
	\label{fig:Gru_1}
\end{figure}

With respect to the RNN, Figure \ref{fig:Gru_0} shows that for low noise levels the correlation is very high regardless of frequency. This is in agreement with the low $J$ shown in Table \ref{tab:Comparison}. Recall the low number of parameters of the RNN with respect to the CNN. However, as the white noise and frequency increase, the number of captured pulses and the accuracy drop more sharply than in the CNN. Besides, when the noise is increased sometimes the training of the RNN does not converge appropriately yielding poor results (see middle left panel of Figure \ref{fig:Gru_0}). Therefore, we can conclude that the proposed RNN has very high performance but only for low levels of noise and pulse frequency. Curiously, for Brownian noise, the performance is similar to that of the CNN (Figure \ref{fig:Gru_1}). This could be explained by the fact that the RNN takes into account specifically the last part of the sequence of the input signal, which is also how is generated brownian noise. 

\begin{figure}[!ht]
	\centering
	\includegraphics[scale=0.65]{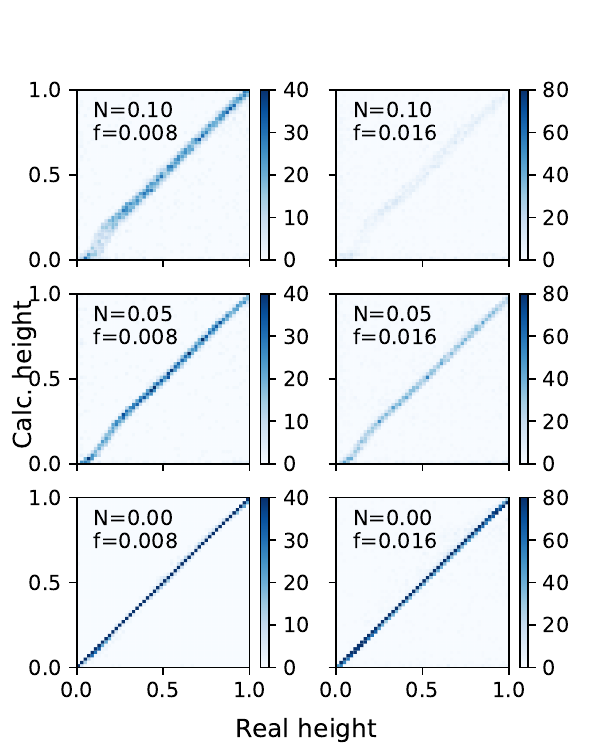}
	\caption{Relationship between the real pulse height and the calculated using the proposed self-attention NN for different levels of white noise $N$ and pulse frequency $f$.}
	\label{fig:Selfatt64_0}
\end{figure}

\begin{figure}[!ht]
	\centering
	\includegraphics[scale=0.65]{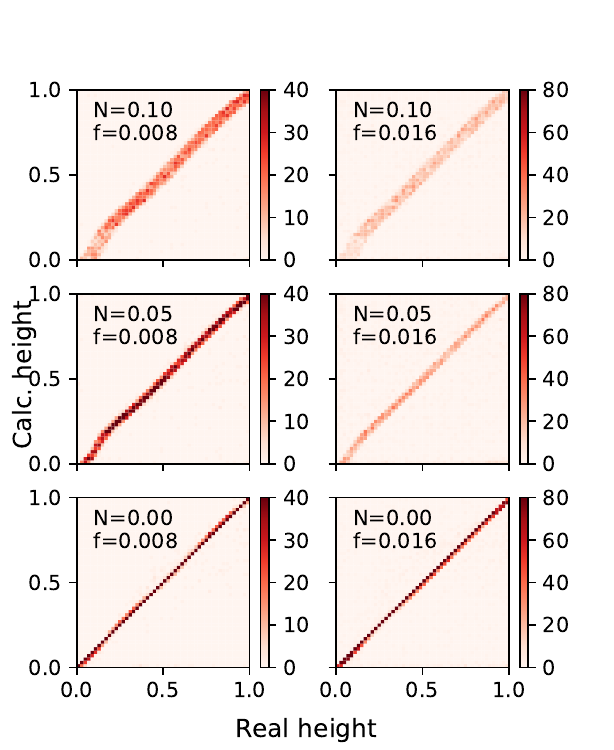}
	\caption{Relationship between the real pulse height and the calculated using the proposed self-attention NN for different levels of brownian noise $N$ and pulse frequency $f$.}
	\label{fig:Selfatt64_1}
\end{figure}

Finally with respect to ANN, Figure \ref{fig:Selfatt64_0} shows that for white noise, the performance is intermediate between CNN and RNN, with high blurring when the noise and frequency are high (see upper right panel of Figure \ref{fig:Selfatt64_0}). With respect to Brownian noise (Figure \ref{fig:Selfatt64_1}) we observe similar performance to the other two NNs. 

Note that ANNs are often most useful in processing signals in several dimensions (e.g. images) because they can identify specific regions and focus their processing on them as pointed out in Section \ref{Introduction}. In contrast, in a one-dimensional signal, there are no spatial regions to focus on, which limits the capability of attentional layers. In fact, the region where ANNs are most focused for pulse height extraction is always in the last input values. Anyway, these last two figures leads us to conclude that the attention mechanism (or the increase in the number of parameters (see Table \ref{tab:Comparison})) does produce an increase in performance with respect to the RNN which, although it does not reach that of the CNN, can be acceptable for certain types of detectors, saving the number of parameters and computational capacity.

The proposed NNs are non-linear systems, so the noise equations to calculate the noise impact or the Equivalent Noise Charge (ENC) \cite{Goulding1972, Workman2022, Regadio2016} cannot be applied to compare them with other linear filters such as FIR or IIR. However, an alternative method to evaluate the NN is to compare the Full Width at Half Maximum (FWHM). In \cite{Regadio2021} it is shown that the use of a U-net type NN such as the one we have used in Section \ref{Convolutional NN} produces significantly better results than a FIR filter (in this case of order 5, whose transfer function was $h(t) = \frac{1}{5}(1,1,1,1,1,1)$, that is a low-pass filter) and linear unfolders as presented in \cite{Jordanov2016, Jordanov2022} or in \cite{Regadio2018}. These statements are also in agreement with what was shown in \cite{Wang2022}, when shaping the signal. We believe that nonlinearity gives a fundamental advantage to nonlinear shapers over classical shapers.

%%%%%%%%%%%%%%%%%%%%%%%%%%%%%%%%%%%%%%%%%%%%%%%%%%%%%%%%%%%%%%%%%%%%%%%%%%%%%%%

\section{Conclusions}
\label{Conclusions}

We have presented three NN topologies that unfold the incoming pulses from energetic particle detectors, return their height, correct the pile-up and even estimate the height of the pulses in presence of saturation. This is possible due to the non-linearity presented by filters based on NNs. The architectures presented are more flexible than linear shapers and can be configured according to the user needs. Additionally, the cost function proposed in this paper, specifically defined for pulse unfolding, significantly improves the results of the three NNs, especially the RNN. On the whole, these NNs provides a better performance compared to more traditional shaping methods at high pulse frequency and noise. We conclude that when noise and frequency are low, all the topologies presented are a valid solution. In the presence of high white noise and pulse arrival frequency, CNN is a better solution than the others. In case of brownian noise, the three presented topologies give similar results. The proposed ANN is an intermediate solution between the CNN and the RNN, since the attention mechanism has limited effectiveness when it comes to unfold the pulses. Finally, these NNs can be trained in a time of the order of minutes; therefore, they can be used for real analysis of pulses coming from particle detectors.

%%%%%%%%%%%%%%%%%%%%%%%%%%%%%%%%%%%%%%%%%%%%%%%%%%%%%%%%%%%%%%%%%%%%%%%%%%%%%%%

\section*{Acknowledgements}

Thanks to project SBPLY/21/180501/000170, cofunded by the Junta de Comunidades de Castilla-La Mancha and the Programa Operativo FEDER (Fondo Europeo de Desarrollo Regional).

\bibliography{thebib}

\end{document}